\documentclass[conference,compsoc]{IEEEtran}
\usepackage{multirow}
\usepackage{gensymb}
\usepackage{makecell} 
\usepackage[table]{xcolor} 
\usepackage{gensymb} 
\usepackage{array}            
\usepackage{enumitem}         
\usepackage{longtable}        
\usepackage{lipsum}
\usepackage{booktabs}
 \usepackage{url}
 \usepackage{rotating}

\sffamily

\usepackage{amsfonts}
\usepackage{amssymb}
\usepackage{ragged2e}

\usepackage{tikz}
\usetikzlibrary{shapes.geometric, arrows}

\tikzstyle{startstop} = [rectangle, rounded corners, 
minimum width=3cm, 
minimum height=1cm,
text centered, 
draw=black, 
fill=red!30]

\tikzstyle{io} = [trapezium, 
trapezium stretches=true, 
trapezium left angle=70, 
trapezium right angle=110, 
minimum width=3cm, 
minimum height=1cm, text centered, 
draw=black, fill=blue!30]

\tikzstyle{process} = [rectangle, 
minimum width=3cm, 
minimum height=1cm, 
text centered, 
text width=3cm, 
draw=black, 
fill=orange!30]

\tikzstyle{decision} = [diamond, 
minimum width=3cm, 
minimum height=1cm, 
text centered, 
draw=black, 
fill=green!30]
\tikzstyle{arrow} = [thick,->,>=stealth]
\usepackage{graphicx}

\usepackage{xcolor}

\usepackage{tikz}

\usepackage{circuitikz}

\usetikzlibrary{positioning}  

\ifCLASSOPTIONcompsoc
  \usepackage[nocompress]{cite}
\else
  \usepackage{cite}
\fi

\ifCLASSINFOpdf
\else

\fi

\usepackage{dirtree}
\usepackage{xcolor}
\usepackage{algorithm}
\usepackage{algpseudocode}
\usepackage{graphicx}

\usepackage[margin=0.5in]{geometry}
\usepackage{textcomp}
\usepackage{pgfplots}
\pgfplotsset{width=8.5cm,compat=1.9}

\begin{document}

\title{Cyber security of Mega Events: A Case Study of Securing the Digital Infrastructure for MahaKumbh 2025 -- A 45 days Mega Event of 600 Million Footfalls}

\author{%
\IEEEauthorblockN{Rohit Negi, Amit Negi, Sandeep K. Shukla}
\IEEEauthorblockA{CSE, IITK}

\and
\IEEEauthorblockN{Manish Sharma}
\IEEEauthorblockA{C3ihub, IITK}

\and
\IEEEauthorblockN{S. Venkatesan}
\IEEEauthorblockA{CSE, IIITA}

\and
\IEEEauthorblockN{Prem Kumar}
\IEEEauthorblockA{IG, UP Police}

}

\maketitle

\begin{abstract}
Mega events such as the Olympics, World Cup tournaments, G-20 Summit, religious events such as MahaKumbh are increasingly digitalized. From event ticketing, vendor booth or lodging reservations, sanitation, event scheduling, customer service, crime reporting, media streaming and messaging on digital display boards, surveillance, crowd control, traffic control and many other services are based on mobile and web applications, wired and wireless networking, network of Closed-Circuit Television (CCTV) cameras, specialized control room with network and video-feed monitoring. Consequently, cyber threats directed at such digital infrastructure are common. Starting from hobby hackers, hacktivists, cyber crime gangs, to the nation state actors, all target such infrastructure to unleash chaos on an otherwise smooth operation, and often the cyber threat actors attempt to embarrass the organizing country or the organizers. Unlike long-standing organizations such as a corporate or a government department, the infrastructure of mega-events is temporary, constructed over a short time span in expediency, and often shortcuts are taken to make the deadline for the event. As a result, securing such an elaborate yet temporary infrastructure requires a different approach than securing a standard organizational digital infrastructure. In this paper, we describe our approach to securing MahaKumbh 2025, a 600 million footfall event for 45 days in Prayagraj, India, as a cyber security assessment and risk management oversight team.  We chronicle the scope,  process, methodology, and outcome of our team's effort to secure this mega event. It should be noted that none of the cyber attacks during the 45-day event was successful. Our goal is to put on record the methodology and discuss what we would do differently in case we work on similar future mega event. We believe that this experience on record could help other similar mega-event organizers in the future in securing their cyber infrastructure. 
\end{abstract}


\IEEEpeerreviewmaketitle

\section{Introduction}
Most cyber security professionals are used to working for organizations where the digital and networking infrastructure is long-lived and has evolved over a long period of time. Most organizations have documented cyber security policy and well-defined processes, have proper asset inventory, vulnerability management tools, configuration benchmarks, well-defined patch management process, well established perimeter security and VPN and MDM based security of remote work arrangement, end-point detection and response tools, network monitoring, security event management and incident response mechanism, backup and restoring process, etc. Audit and vulnerability assessment and penetration testing exercises are also a routine process for most well-governed organizations. Imagine such professionals encountering a temporary organization which is elaborately digitalized in terms of wired and wireless network, a multitude of web and mobile application based services, CCTV-based  surveillance, AI based crowd management system, elaborate control center with 24x7 surveillance of video feed, network and end-point monitoring, and where the entire digital infrastructure is built over a few weeks or months, and would be dismantled at the completion of a mega event.
Furthermore, given the cultural and national significance attached to the event, the digital infrastructure is the target of cyber threat actors. How do we ensure that the necessary cyber security framework has been put in place? The National Institute of Standards and Technology (NIST) cyber security framework \cite{NIST2024CSF} requires six functions to be implemented, namely {\em govern, identify, protect, detect, respond and recover}. However, given the compressed time frame, the temporary nature of the infrastructure, the dependence of multiple software vendors and their data centers, and the lack of awareness and training among the personnel involved,  it is quite a difficult challenge, as our team experienced.   

Although our experience chronicled in this paper is related to a religious mega event, the same lessons apply to other mega events such as sports tournaments, global expositions, diplomatic summits, etc., as these are all complex digital ecosystems where organizers intend to demonstrate technological innovation and cultural strength, and the infrastructure is temporary.   Mega events involve millions of participants and happen over a short period of time (few days to several weeks duration events), require creation of temporary infrastructure including not only the infrastructure for the main event, but also peripheral infrastructure such as ticketing,  food and beverage, shops, sanitation facilities, control rooms, lost-and-found facilities, surveillance system for public safety and crowd control, public address systems, digital display systems, transportation facilities, waiting rooms, etc.

 Mega events these days are highly dependent on digital technologies and Information and Communication Technology (ICT) infrastructure.  Although these innovations can increase public safety and operational efficiency, they also expand the attack surface for malicious actors targeting such events. 

In the recent past, we have seen multiple such cyber attack incidents during mega events. For example,  \cite{clay2024_trendmicro}, \cite{govtech_qotd2021}, \cite{greenberg2019untold},  \cite{microsoft-cyber-signals5}, \cite{singh2024_hindu} ,\cite{socradar2022_phishing}, \cite{tan2021_tokyo_data} (see Table \ref{tab:megaevent-cyber-incidents}) describe cyber incidents in Tokyo Olympics 2020, Winter Olympics in South Korea in 2018, FIFA World Cup in 2022, G-20 Summit in India, etc. Such information available in the public domain is sufficient to believe that mega-events are becoming lucrative targets for cyber attacks. Cyber security failures can cause reputational damage, financial loss, and erosion of public trust. Sometimes, mega-events process high volumes of personal and operational data. Such data storage also becomes attractive targets for cyber attackers to commit downstream cyber crime and espionage.

\begin{table}[htbp]
  \caption{Notable Megaevent Cyber Incidents}
  \label{tab:megaevent-cyber-incidents}
  \centering
  \begin{tabular}{>{\raggedright\arraybackslash}p{2cm} c >{\raggedright\arraybackslash}p{2cm} >{\raggedright\arraybackslash}p{2.8cm}}
    \toprule
    Event & Year & Type of Attack & Impact \\
    \midrule
    Winter Olympics, Pyeongchang & 2018 & Olympic Destroyer malware & Disrupted opening ceremony, Wi‑Fi, ticketing \\
    \addlinespace[6pt] 
    Tokyo Olympics & 2020 & 450 M attempted attacks, data breach & Minor breach, no operational impact \\
    \addlinespace[6pt]
    Super Bowl (49ers) & 2022 & Ransomware & Team systems affected, unclear event impact \\
    \addlinespace[6pt]
    FIFA World Cup, Qatar & 2022 & Phishing, fake apps, DDoS plans & Threats mitigated, no major disruptions \\
    \addlinespace[6pt]
    G20 Summit, India & 2023 & DDoS, targeted attacks & Attacks thwarted, no operational impact \\
    \bottomrule
  \end{tabular}
\end{table}

Interestingly, this year, the Mahakumbh 2025 \cite{pib2025mahakumbh}, \cite{pib2025mahakumbhbegin} event was also declared as a {\em digital} Mahakumbh 2025 \cite{pib2024digitalmahakumbh}, \cite{2090956} by the prime minister of India, given that India is pushing for an inclusive and universal digitalization in all facets of citizen's lives including digital services to citizens by the government, digital payment infrastructure \cite{2103221} such as Unified Payments Interface (UPI),  digital identity infrastructure, and digital India stack as a public good. 

This Mahakumbh event was projected to attract more than 500 million pilgrims (eventually it exceeded 600 million)  with a massive gathering on certain special days, characterized by profound spiritual significance and immense logistical demands. 
Public safety and smooth operation depended on secure, scalable, and resilient digital systems for crowd management, health surveillance, cashless transactions, and emergency response operations using critical infrastructure: pontoon bridge, power grids, transport systems, and water supplies, often managed through connected systems like SCADA and IoT. The venue became a temporary city for 45 days with all utilities available to devotees and the venue has its own local government with a district magistrate at the helm of the administration of the temporary district. 

The unique blend, that is, the usage of deep tech and participation from different backgrounds, made the event incredibly exciting, but also complex from the organizer's point of view. Ensuring a seamless and engaging experience for all participants while maintaining robust cyber security becomes both a challenge and a responsibility.

Our focus in this paper is to chronicle the challenges we faced during our continuous engagement with the organizing administrative team and technology partners involved in the event to ensure that a cyber security framework is in place to provide cyber resilience in the event of a large-scale cyber attack and cyber protection from smaller-scale cyber incidents. Since the digital and network infrastructure was evolving from one month before the start of the event until the event started in full swing, our challenges were quite distinct from our prior experience with long-standing organizations. We also discuss and demonstrate that while our overall approach was structured in the NIST Cyber Security Framework (NIST CSF 2.0) \cite{NIST2024CSF}. Our approach to auditing cyber security controls was inspired by the ISO/IEC 27001 \cite{ISO27001} and ISO/IEC 27002:2022 \cite{ISO27002} standards. Since a mega-event is not a permanent infrastructure, and time limited temporary in nature, a full blown ISO 27001 approach could not be taken as the engagement started only about 40 days before the event start date, but our approach tried to enforce the ISO 27001 control regimen on the organizations involved. To the best of our knowledge, no prior documented approach to mega-event  organization is currently available in the cyber-security literature. Based on our experience, we also provide certain recommendations for future organizers of such mega-events.  The main contributions of this paper are:

\begin{enumerate}
    \item Documenting the challenges of developing a cyber security framework to create cyber resilience for a megaevent characterized by its temporariness and the expedient development of ICT infrastructure. 

    \item Demonstrating the application of NIST CSF 2.0 \cite{NIST2024CSF} to achieve cyber resilience for highly digitalized megaevents in a short period of time prior to the start of the event. 
    
    \item Demonstrating an ISO/IEC 27001/27002:2022 \cite{ISO27001}, \cite{ISO27002} inspired baseline control list, we developed to check the security posture of the vendors providing ICT infrastructure and the applications. 
    
    \item Developing and chronicling a cyber security methodology for short-duration mega-events to aid the decision-making of event organizers.
\end{enumerate}

\subsection{Organization}
In Section 2, we discuss related work on the cyber security of mega-events. In Section 3, explain the problem statement, particularly the problem of achieving resiliency of the event organization through achieving cyber resilience. In Section 4, we detail the scope of the digitalization of the  Mahakhumbh event. In Section 5, we detail our methodology to achieve cyber resiliency. Section 6 concludes the paper with some discussions on how to improve the process in the future.

\section{Background \& Related work}

In the modern world, ICT infrastructure becomes an integral part of mega events, as shown by a plethora of case studies in the ICT literature.  For example, in \cite{6957980}, a ``shiftable'' intelligent transportation system (ITS) was shown to be effective in arranging traffic control during a traffic surge due to a mega event organized in an otherwise low-traffic region.  It was also shown that it reduces the cost of transportation by 96\% over the conventional intelligent transportation system. In \cite{elkhouly2023mitigating}, an agent-based simulation case study is presented to evaluate the use of innovative ICT infrastructure to enable crowd management in metro, bus and train stations during crowd surge caused by mega events.  In \cite{santomier20162012},  a case study of the London Olympics mega event in 2012 has been detailed. The London organizing committee of the Olympic and Paralympic Games (LOCOG) and the Olympic Delivery Authority (ODA) used ICT infrastructure and social media, the Olympic Commentator Information System (CIS) and the Information Diffusion System (IDS) for better management of the mega-sport event.  In a proposed \cite{garay2014state}  integration of services in preparation for the FIFA World Cup and Olympics in Sao Paolo, Brazil, it was proposed to integrate services such as Telecommunications, Urban Mobility, Airports, Tourism, Energy and Security with the city infrastructure in preparation for two mega-events. 

\begin{figure}[!htb]
    \centering
    \includegraphics[scale=0.5, trim={0.5cm 0 0 0}, clip]{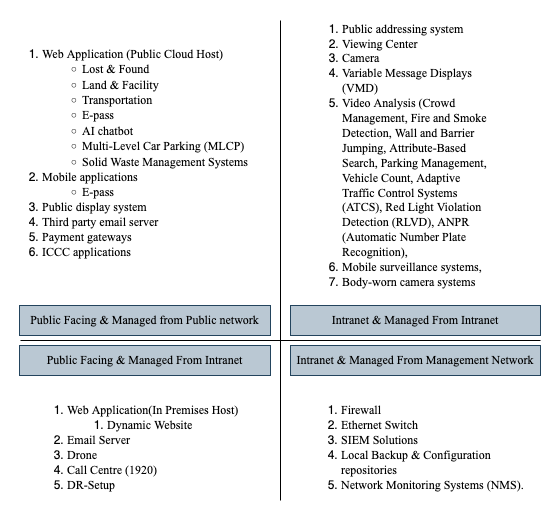}
    \caption{Major ICT infrastructure used during the execution of Digital Mahakhumbh}
    \label{fig:ict}
\end{figure}

As depicted in Figure ~\ref{fig:ict}, Digital Mahakhumbh \cite{2088994} has been carried out with the support of ICT infrastructure such as 360\degree ~virtual reality stalls, spectacular drone shows, web and mobile applications for land and facility allocation, sanitation services, lost and found, reservation of transportation and tents, remote controlled life bouys, underwater drones, AI powered cameras, etc.

In the recent past, such extensive use of the ICT infrastructure to facilitate mega events has made them potential targets of cyber threat actors. In particular, nation-state threat groups such as APT groups, cyber criminal gangs, as well as hacktivists and hobby hackers have been known to target various parts of the infrastructure to induce large-scale disruption and panic, showing off their hacking capability, committing financial frauds, even indirectly inducing stampede and other forms of chaos, leading to the embarrassment for the organizers or organizing countries, etc. 

Several cybersecurity incidents have been reported that affected various mega events, such as the 2018 Winter Olympics in Pyongyang \cite{greenberg2019untold}.  An intrusion into the internal networks of the host team that exploited latent vulnerabilities, strained communication, and incident response resources. In contrast, the Tokyo 2020 Olympics \cite{govtech_qotd2021} demonstrated the effectiveness of robust cyber security measures taken before the event: More than 220 ethical hackers, trained by the Japan National Institute of Information and Communications Technology, helped prevent potential disruption despite millions of attempted intrusions \cite{tan2021_tokyo_data}. In the case of Super Bowl 2022 (San Francisco 49ers) \cite{microsoft-cyber-signals5}, ransomware and domain controller compromises led to stadium WiFi failure, blackout of Internet-linked televisions, disruption of RFID-enabled security gates, and breakdown of the official ticketing system, necessitating the deployment of WiFi hot spots (Wireless Fidelity) and manual badge checks to restore operations.  Similarly, the FIFA World Cup 2022 \cite{socradar2022_phishing} in Qatar faced intense threat activity, including phishing campaigns, fake ticketing apps, and chatter on the dark Web. Fortunately, they did not experience major operational impacts, thanks to a multilayered cyber
security defense strategy \cite{clay2024_trendmicro}. At the G20 Summit of 2023 \cite{singh2024_hindu} in India, coordinated DDOS and targeted intrusion attempts were successfully neutralized by Indian cyber security agencies, including CERT-In (Indian Computer Emergency Response Team) and I4C (Indian Cyber Crime Coordination Center). It is now mandatory that cyber experts conduct an evaluation of the mega-event ICT setup and consider remediation measures to reduce the impact of cyber attacks on the mega-event setup.

In \cite{seloom2024qatar} it is mentioned that during the 2022 FIFA World Cup, approximately 49,000 security personnel were deployed during the event in Qatar to present an advanced model of security management for global mega-events. However, it is not clear how many of these personnel were used for cyber-security operations. Most of the existing literature \cite{bongiovanni2024protecting}  on security and sports events focuses on physical security. However, as mentioned earlier, mega sports events have been under increasing threat of cyber attacks in recent years. Cyber security in mega-events requires a structured, standardized approach leveraging internationally recognized frameworks. There is a scarcity of literature in the area of cyber security management for mega-events.

It has been pointed out in the previous section that for established and long-lived organizations, the practices of assessing cyber threats, vulnerabilities, cyber risks, control adequacy, secure configuration settings, timeliness of patch management, efficacy of security operations and processes are well understood. 
Several frameworks and guidelines are available \cite{hamdani2021cybersecurity}. Among many other documents in the literature, \cite{djebbar2023comparative}, \cite{hamdani2021cybersecurity} are examples of work where a detailed analysis of cyber security frameworks and standards such as The European Telecommunications Standards Institute (ETSI) 303 645 \cite{ETSI2024EN303645},  National Institute of Standards and Technology (NIST) \cite{SP80053Ar5}, \cite{NIST2020SP500-332}, \cite{NIST2020SP800-53r5}, \cite{NIST2018SP800-37r2}, Federal Information Processing Standards (FIPS) \cite{nistfips},  ISA/IEC 62443 \cite{isa62443},  Common Criteria (CC-Cyber Security) \cite{commoncriteria}, \cite{fekete2011common}, and ISO/IEC 27001/2 \cite{ISO27001}, \cite{ISO27002} is provided. In \cite{dimakopoulou2024comprehensive}, NIST CSF \cite{NIST2024CSF} has been used for a comprehensive analysis of cyber security of maritime organizations, and it revealed significant gaps in the academic and industry-specific literature. In \cite{tsiodra2023cyber}, CIS (Center for Internet Security) Controls \cite{ciscontrols} benchmarks and the CWE (Common Weakness Enumeration)  \cite{cwe} top 25 checklist have been used to conduct the assessment using a sample network topology of a small business. A local government security review using NIST CSF in \cite{ibrahim2018security} demonstrated the adoption of the framework as an assessment toolkit that targets different levels of the organization. In addition to this, a skill set assessment methodology \cite{budde2023consolidating} is also available to implement people control. None of these applies to temporary yet very large ICT enabled events where the infrastructure is built in a short time span, and the usage of the infrastructure is for a limited time duration beyond which the entire system is dismantled.  While working on cyber security preparedness and assessment of Mahakumbh 2025, we realized this knowledge gap, and this work is our attempt to fill that gap to some extent. 

Vulnerability assessment and penetration testing (VAPT) \cite{6724216}, \cite{10094095}, \cite{7583912}, \cite{8862224}, \cite{10891284}, \cite{9511495}, \cite{10111168}, \cite{negi2019vulnerability} are often conflated with cyber security posture assessment of the digital infrastructure created for a mega event. Such a misguided approach might remove high-severity vulnerabilities in the applications but it fails miserably in providing cyber resiliency. Although VAPT is necessary to identify vulnerabilities in the various applications, network, devices used, and the severity of the identified vulnerabilities in terms of CVSS scoring \cite{first2023cvss} helps prioritize urgency of remediation, it is not sufficient to simply perform VAPT and fix vulnerabilities found. This paper captures our approach to securing a mega-event, along with our recommendations for future mega-events, and our approach and recommendations go far beyond VAPT. 

A more common approach to verifying the cyber security preparedness of a mega event would be to carry out a compliance audit. A compliance audit is an assessment of the information security management system (ISMS) \cite{broderick2006isms}, \cite{asosheh2013practical}, \cite{haufe2016isms}, \cite{chavez2024implementation} according to an ISMS standard such as ISO 27001.  It provides a structured way of conducting the cyber security assessment of the organization -- in case of mega events -- a temporary organization. The latest version of ISO/IEC 27001: 2022 prescribes 93 controls, which are further categorized as organizational, people, physical, and technological. During the assessment using an ISMS road map, there is the concept of a Statement of Applicability (SOA). In SOA, an auditor first identifies the applicability of the controls to the organizations based on which audit is conducted.  In case a few controls  are exempted,  a justification based on legal, business, or regulatory requirements must be documented and agreed upon.

As part of preparatory steps, we investigated the existing cyber security literature related to securing mega events from the recent past. Interestingly, almost all of the literature, opinion pieces, blogs, research articles, and white papers were related to mega sports events such as the Olympics, FIFA World Cup, and similar large-scale sporting tournaments, or diplomatic meetings such as the G20 Summit. We could not locate one related to a cultural mega event like Mahakumbh and we found no comprehensive article chronicling the cyber security preparedness of mega events organized in India even though a number of such events recently took place in India. It is therefore imperative for us not only to describe what, why, when, where and how of our preparedness, but also to provide our recommendations based on the learning of the experience. 

In \cite{bongiovanni2024protecting}, a comprehensive review of the literature can be found on the cyber security posture of sports events. The article finds that not enough articles have been published on this very important topic even though a number of high-profile attacks have taken place on recent mega sports events. The article also observed that only 28 relevant papers were found, of which many of them discuss physical security as well as cyber security, as an overall security issue. In addition, they found no agreed upon cyber security framework emerging out of the literature. In \cite{cyberdefense2025}, \cite{ctareport2024} \cite{fowler2024} and \cite{duxbury2024} are popular articles that discuss the acute need for cyber security to protect highly ICT-integrated sports events, but do not provide any framework or methodology prescription. On the other hand, \cite{cltc2017} discusses cyber risks in great detail, in major sports events such as the Olympics, in the context of the opportunities of digitalization-based efficiency of running such events. In addition, \cite{talos2024} is a vendor-specific white paper but provides 13 strategies to improve the cyber security posture of mega sports events. These include {\em asset discovery and threat modeling; intrusion detection and network segregation; remote access security and monitoring;, awareness training of personnel; endpoint protection and hardening; data storage and access control; incident response planning; risk management; backup and recovery procedure planning and drills, threat intelligence and SOAR; vendor management and event security management; unified ticketing platform with 24x7 support;} and {\em proactive threat hunting}. Although this is a comprehensive list that helped our approach, it is not a framework to serve as a scaffold to our approach for Mahakumbh 2025.  Finally, another book draft \cite{whelan2018} is helpful, but it does not focus on cyber security and dwells on the general security of mega-events. 

Digital Mahakumbh 2025 turned out to be one of the largest cultural congregations worldwide.  The lessons from past cyber incursions during mega events formed a crucial basis for pre-emptive planning.  In order to check security posture of the Mahakhumbh's infrastructure, we took inspiration from the ISO / IEC 27001: 2022 \cite{ISO27001} and ISO / IEC 27002: 2022 \cite{ISO27002} standards, which establish foundational practices through systematic Information Security Management Systems (ISMS), enabling event organizers to identify, assess, and mitigate cyber security risks comprehensively, from physical access control to network encryption and incident management. However, a NIST Cyber Security Framework (CSF 2.0) \cite{NIST2024CSF}-based approach informed our overall functioning as the team that oversaw the cyber security of the event.

\section{Problem Statement: Mission Resiliency with Cyber Resiliency}
In order to fully appreciate the problem at hand, we must recall that the {\em mission} of the Mahakumbh organizers was to provide a seamless pilgrimage experience for more than 500 million pilgrims (projected to be 500 million but the actual figure turned out to be 660 million) from all over India and outside of India, gathered in the city of Prayagraj at the precise location, which is marked by the meeting point of three rivers,  over a short period spanning 45 days from January 13, 2025 to February 26, 2025.  In addition,  six specific dates within this period were declared particularly important, and the number of pilgrims on those days was projected to exceed 20 million each time. 

In addition, this year, Mahakumbh was also declared Digital Kumbh due to the extensive use of digital technology \cite{pib2024digitalmahakumbh}, software applications, crowd control enabled by AI, e-pass, and other artifacts based on mobile devices. Therefore, although periodic garbage collection, sanitation, availability of health facilities on site, physical security, crowd management, transportation arrangement, food arrangements, vendor management, etc., were important to achieve the overall mission, without achieving {\em cyber resiliency}  mission resiliency would not have been possible. Since most of the business processes were ICT enabled, the overall resilience of the mission depended on cyber resilience. 
This brings us to the problem statement presented to our team about a month before the start of the event.  Our team was asked to oversee the cyber security of the event, which we explained to the authorities that must extend to an oversight of cyber resiliency. 

According to \cite{nistsp800160v2}, cyber resilience is defined as follows: 
\begin{quote}
    ``\textit{Cyber resilience is the ability to {\bf anticipate, withstand, recover from}, and {\bf adapt to adverse conditions, stresses, attacks, or compromises} in systems that use or are enabled by cyber resources}''.

    A cyber resource is an information resource that creates, stores, processes, manages, transmits, or disposes of information in electronic form and that can be accessed through a network or by using networking methods.
\end{quote}

 The problem our team was tasked with solving was to assess the cyber resiliency of the digital infrastructure of Mahakumbh 2025 and recommend the various processes and cyber security controls to enhance the resiliency, until it reached an acceptable level of resiliency. However, our team had to define the acceptable level and if the level of resiliency is not achieved to our satisfaction before the start date, the organizers had to take some strong decisions regarding the continued use various applications and digital processes. Furthermore,   the cyber crisis management plan, the incident response plan, and the business continuity plans had to also be included in our recommendations. 

 During a forty-day period prior to the start of the mega event, our team met with the administrative heads of the organizing team both physically on site and remotely via video conferencing frequently. The frequency of the meetings increased rapidly as the date approached. In each subsequent meeting, risk reduction was discussed, vulnerability patching progress was discussed with vendors, and our own assessments, threat models, mitigating control efficacy, and suggested administrative control in the event of inadequacy of controls were thoroughly discussed with the organizing team. 


\section{The Scope: Digital Components of the Event}
As mentioned earlier, the organizers wanted to create a seamless and hassle-free experience for vendors, pilgrims, law enforcement personnel, organizers, and event organizers for the various concurrent events organized with the big Mahakumbh event. The goal was to use ICT technology, combined with IoT devices such as CCTV cameras, gates sensors, wireless and wired network-based communication, cyber security technology, web and mobile applications to deliver the best of class services.

\begin{table}[h]
    \centering
    \caption{Service Providers and Key Security Aspects Assessed}
    \label{tab:service_providers}
    \begin{tabular}{>{\raggedright\arraybackslash}p{1.2cm} >{\raggedright\arraybackslash}p{0.3\columnwidth} >{\raggedright\arraybackslash}p{4cm}}
        \toprule
        \textbf{Vendor} & \textbf{Services Provided} & \textbf{Key Security Aspects Assessed} \\
        \midrule
        \multirow{3}{*}{$V_{1}$}  & CCTV & Device \& Network security. \\
        & Monitoring Service & Encryption of video feeds. \\
        & VMD & Access control, data storage protection. \\
        \addlinespace[6pt]
        \multirow{3}{*}{$V_{2}$}  & Police Vehicle Tracking System & GPS tracking security. \\
        & PAS & Endpoint security. \\
        & Body Worn Camera & Firmware integrity. \\
        \addlinespace[6pt]
        \multirow{3}{*}{$V_{3}$} & CCTV & Device authentication. \\
        & Monitoring Service & Data transmission security. \\
        & PAS & Risk assessment of VMD display hacking. \\
        \addlinespace[6pt]
        \bottomrule
    \end{tabular}
\end{table}

Infrastructure services deployed in Prayagraj City and the Mahakumbh area included critical surveillance and communication technologies such as CCTV cameras, variable message displays (VMD), public announcement systems (PAS) and police vehicle tracking systems. 
Three different vendors provided infrastructure services. The services provided by these vendors and the main security aspects evaluated are indicated in Table ~\ref{tab:service_providers} (the exact identities of the vendors are anonymized). 

\begin{table}[H]
    \centering
    \caption{Vendor role and responsibility}
    \label{tab:my_label}
    \begin{tabular}{p{1cm} | p{1.5cm} p{2.5cm} p{2.5cm}}\hline
     Vendor    & Infra/PMU & Web Application & Mobile Application  \\\hline
     $V_{1}$   & 1 & - &  - \\
     $V_{2}$   & 1 & - & - \\
     $V_{3}$   & 1 & - & - \\
     $V_{4}$   & - & 2 & 3 \\
     $V_{5}$   & - & 2 & - \\
     $V_{6}$   & - & 1 & - \\
     $V_{7}$   & - & 1 & - \\
     $V_{8}$   & - & 1 & - \\
     $V_{9}$   & - & 1 & - \\
     $V_{10}$  & 1 & 2 & 3 \\
     $V_{11}$  & - & 1 & 1 \\
     $V_{12}$  & - & 1 & - \\\hline
    \end{tabular}

\end{table}

\begin{table}[h]
    \centering
    \caption{Web Applications Assessed In-Scope}
    \label{tab:web_applications}
    \setlength{\extrarowheight}{4pt} 
    \begin{tabular}{>{\raggedright\arraybackslash}p{0.5cm} >{\raggedright\arraybackslash}p{0.3\columnwidth} >{\raggedright\arraybackslash}p{4.7cm}}
        \toprule
        \textbf{\#} & \textbf{Web Application} & \textbf{Purpose / Functionality} \\
        \midrule
        1 & Mahakumbh Website & Information portal for pilgrims and visitors. \\
        2 & Kumbh Sah`AI'yak Chatbot & AI-powered virtual assistant for query resolution. \\
        3 & UP State Tourism Website & Tourism information and booking services. \\
        4 & Digital Lost \& Found Center & Reporting and recovering lost items. \\
        5 & Land \& Suvidha Allotment Portal & Tent and facility allotment management. \\
        6 & ICT Monitoring of Sanitation \& Tentage & Real-time monitoring of cleanliness and tent conditions. \\
        7 & PDS Automation System & Public Distribution System tracking and automation. \\
        8 & FRB System & Facial Recognition Base system for emergencies. \\
        9 & Inventory Tracking System & Real-time tracking and management of assets. \\
        10 & Digital Display System & Information boards and digital announcements. \\
        11 & Kumbh Website for Officials Only & Restricted access site for event authorities. \\
        12 & E-Pass Portal & Digital entry pass system for authentication. \\
        \bottomrule
    \end{tabular}
\end{table}

A total of 12 web and 7 mobile applications were developed by 9 distinct vendors. These applications were developed specifically for Mahakumbh 2025 to assist pilgrims with tent allotment, e-passes, ration distribution, incident reporting, information dissemination, etc. The list of web applications is provided in Table ~\ref{tab:web_applications} and a separate list of mobile applications is provided in Table ~\ref{tab:mobile_apps}. 

\begin{table}[h]
    \centering
    \caption{Mobile Applications Assessed In-Scope}
    \label{tab:mobile_apps}
    \begin{tabular}{@{} c >{\raggedright\arraybackslash}p{0.25\columnwidth} >{\raggedright\arraybackslash}p{0.15\columnwidth} >{\raggedright\arraybackslash}p{0.45\columnwidth} @{}}
        \toprule
        \textbf{\#} & \textbf{Mobile Application} & \textbf{Platform} & \textbf{Purpose / Functionality} \\
        \midrule
        1 & Maha Kumbh Mela App & iOS \& Android & Provides public information and essential services for pilgrims. \\
        \addlinespace[6pt] 
        2 & E-Pass Portal & iOS \& Android & Digital entry management system for event attendees. \\
        \addlinespace[6pt] 
        3 & Kumbh Rail Sewa & Android & Railway service tracking for pilgrims and travelers. \\
        \addlinespace[6pt] 
        4 & FRB App & Android & Facial Recognition Base application for emergencies. \\
        \addlinespace[6pt] 
        5 & Kumbh Application for Officials Only & Android & Restricted access application for event coordinators and authorities. \\
        \bottomrule
    \end{tabular}
\end{table}

The scope of our team's work was to assess cyber resiliency of the entire digitalized business processes, cyber security assessment of the infrastructure as well as the applications deployed, and recommending remediation for any shortcoming so that the cyber resiliency is achieved at a significant level by the start date of the event. Note that our team was called to this task about forty days before the start date. 

\section{Methodology for Achieving Cyber Resiliency }

At the beginning, about 6 weeks  before the start date of the mega event, the top administration of the event organization contacted us to serve as a team of subject matter experts in the field of cyber security. The goal of the engagement was not explicitly stated, and most of the discussions were around how to guarantee uninterrupted operations throughout the mega event, avoiding any cyber security mishap. Our understanding after the initial discussion was that the organizers want an assurance from the subject matter experts (SME) that adequate cyber security controls are in place and event is ready to mitigate any cyber threats. It was clear that our methodology should not only be limited to assessing cyber security readiness, but should also play an important role in improving the cyber security posture of the event along with the assurance to the event organizers that all stakeholders are ready to withstand, survive and recover from any unavoidable cyber incident. 

Our team's methodology was guided by NIST 800-160v2 \cite{nistsp800160v2}, the NIST guideline for developing cyber-resilient systems. According to \cite{nistsp800160v2}, a structured approach to engineering systems that can anticipate, withstand, recover, and adapt to cyber threats must include the following key steps:
\begin{itemize}
    \item {\bf Study the Context:} Identify the mission objectives, system dependencies, and threat environment and scope.
    \item {\bf Develop a cyber resilience baseline: } Assess the initial cyber security capabilities, the posture of resiliency, and the cyber security controls in place.
    \item {\bf Assess the Current Cyber Security Posture:  } Find existing vulnerabilities, attack surfaces, and potential tactics and techniques by potential threat actors
    \item {\bf Define and Analyze Alternative Resiliency Techniques: } Identify possible technology, process and techniques and prioritize based on possible choices
    \item {\bf Create and Implement Recommendations: } Implement strategies to enhance resiliency while balancing cost, performance, and risk.
\end{itemize}

For studying the mission and context, members of our team visited the event site for several days and interviewed the organizers, the ICT service providers, software developers, network service providers and the highest echelons of government officials in charge of the event. A summary of the context is already provided in Figure ~\ref{fig:ict}, Table ~\ref{tab:service_providers}, Table ~\ref{tab:web_applications}, and Table ~\ref{tab:mobile_apps}. The assessment of the initial cyber security capabilities, the posture of resiliency, and controls was thoroughly investigated and cataloged by the team through multiple visits, online meetings, and discussions with the various ICT vendors. A detailed report was drafted and shared with the event organizers, and a road map was provided for resiliency engineering. 

Some of the main observations made in the draft report at this point are as follows.
\begin{itemize}
    \item A complete vulnerability assessment and penetration testing for the onsite ICT infrastructure and vendor-provided applications was not exhaustively completed. Table ~\ref{tab:vapt} lists the kind of vulnerabilities discovered during initial VAPT by our team. 
    \item No threat modeling was carried out even though, from previous mega events, certain types of threat were expected by the organizers.  For example, large-scale denial of service attempts, web page defacing attempts, and various types of cyber fraud such as fake websites selling tickets, reservations of tents and vendor stalls, fake websites promising various facilitated services. A proper threat modeling exercise was required. 
    \item In terms of application hosting, the existence of redundant or back up servers, secondary back up servers, etc. were not documented, and for certain vendors it was unclear if they can provide business continuity guarantees if their backend server hosting is affected by a cyber attack. 
    \item A cyber crisis management plan was not documented and no incident response playbooks were documented. 
    \item Due to the commissioning of various services to various vendors, a fragmented end-to-end business continuity plan, a general cyber resilience infrastructure, and vendor/third-party cyber risk management was conceived.   
    \item There was an overall system integration consulting company to coordinate across all vendors, who was our main point of contact to coordinate with all the stakeholders. 
\end{itemize}

The third key step of resiliency engineering, namely, the assessment of current cyber security posture through threat modeling, adversary modeling, vulnerability assessment and penetration testing, risk assessment was carried out until the very end as it took a number of iterations to close the vulnerabilities, estimate residual risks, and get the event authority's assent to the acceptability of residual risks. 

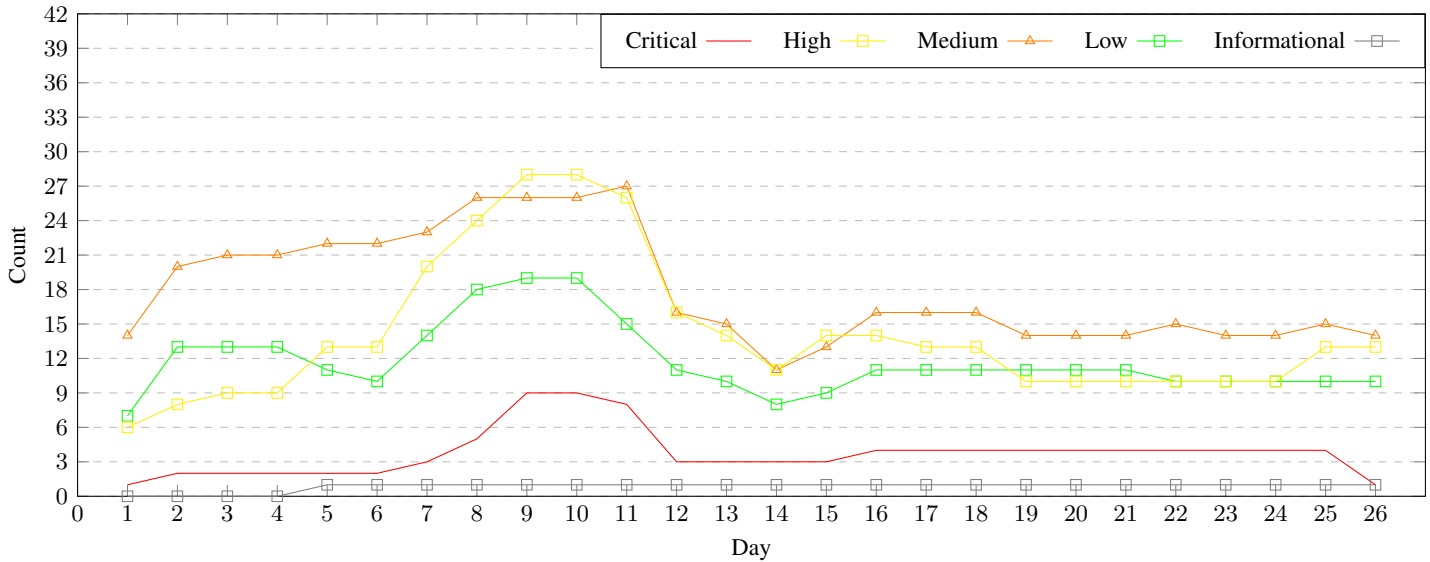
\begin{figure*}[h]
    \centering

    \begin{tikzpicture}[scale=1.0, transform shape]
    \begin{axis}[
        name=plot,
        width=19.5cm,         
        height=8cm,
        xlabel={Day},
        ylabel={Count},
        label style={font=\small},         
        tick label style={font=\small},    
        xmin=0, xmax=27,
        ymin=0, ymax=42,
        xtick={0,1,2,3,4,5,6,7,8,9,10,11,12,13,14,15,16,17,18,19,20,21,22,23,24,25,26},
        ytick={0,3,6,9,12,15,18,21,24,27,30,33,36,39,42},
        ymajorgrids=true,
        grid style=dashed,
    ]

    \addplot[
        color=gray,
        mark=square,
        ]
        coordinates {
        (1,0)(2,0)(3,0)(4,0)(5,1)(6,1)(7,1)(8,1)(9,1)(10,1)(11,1)(12,1)(13,1)(14,1)(15,1)(16,1)(17,1)(18,1)(19,1)(20,1)(21,1)(22,1)(23,1)(24,1) (25,1)(26,1)
        };
        \label{Informational}
        
    \addplot[
        color=green,
        mark=square,
        ]
        coordinates {
        (1,7)(2,13)(3,13)(4,13)(5,11)(6,10)(7,14)(8,18)(9,19)(10,19)(11,15)(12,11)(13,10)(14,8)(15,9)(16,11)(17,11)(18,11)(19,11)(20,11)(21,11)(22,10)(23,10)(24,10) (25,10)(26,10)
        };
        \label{Low}

    \addplot[
        color=orange,
        mark=triangle,
        ]
        coordinates {
        (1,14)(2,20)(3,21)(4,21)(5,22)(6,22)(7,23)(8,26)(9,26)(10,26)(11,27)(12,16)(13,15)(14,11)(15,13)(16,16)(17,16)(18,16)(19,14)(20,14)(21,14)(22,15)(23,14)(24,14) (25,15)(26,14)
        };
        \label{Medium}

    \addplot[
        color=yellow,
        mark=square,
        ]
        coordinates {
        (1,6)(2,8)(3,9)(4,9)(5,13)(6,13)(7,20)(8,24)(9,28)(10,28)(11,26)(12,16)(13,14)(14,11)(15,14)(16,14)(17,13)(18,13)(19,10)(20,10)(21,10)(22,10)(23,10)(24,10) (25,13)(26,13)
        };
        \label{High}
    
    \addplot[
        color=red,
        mark=trianlge,
        ]
        coordinates {
        (1,1)(2,2)(3,2)(4,2)(5,2)(6,2)(7,3)(8,5)(9,9)(10,9)(11,8)(12,3)(13,3)(14,3)(15,3)(16,4)(17,4)(18,4)(19,4)(20,4)(21,4)(22,4)(23,4)(24,4) (25,4)(26,1)
        };
        \label{Critical}
    \end{axis}
    
    \node[anchor=north east,fill=white,draw=black] (legend) at (plot.north east) {\begin{tabular}{lllll}
            \small Critical \ref{Critical} & \small High \ref{High} & \small Medium \ref{Medium} & \small Low \ref{Low}  & \small Informational \ref{Informational}
        \end{tabular}
        };

    \end{tikzpicture}
    \caption{Profiling of Vulnerability Management}
    \label{fig:profiling}
        
\end{figure*}

Our VAPT team performed vulnerability assessment and penetration testing (VAPT) of individual applications and appliances, including CCTV cameras, Visual display boards, onsite wifi network, etc.  The details of the various cyber assets (applications) (see Table ~\ref{tab:web_applications} and Table ~\ref{tab:mobile_apps}) were received gradually in multiple batches. Initially, we received the asset details for nine web applications. The first VAPT report was submitted for 04 of the 09 assets. A total of 28 vulnerabilities were reported (see Table ~\ref{tab:vapt}, and the first column of Table ~\ref{tab:profile}).  The distribution of the types of vulnerability found is listed in Table ~\ref{tab:vapt}. Approximately 25\% of the vulnerabilities found at this stage were related to information disclosure due to configuration issues. Some of the applications were still in continuous integration and deployment mode (CI/CD), and as soon as they made a change and deployed,  newer vulnerabilities were being introduced. This caused us to repeat the VAPT and newer vulnerabilities in the assets were found. Due to CI/CD, we continued to carry out the repeated VAPT until 20 January\textsuperscript{th}, 2025 from 23 December\textsuperscript{rd}, 2024.
 We also performed multiple revalidations on request or confirmation of the mitigation of vulnerabilities by the OEM. Table \ref{tab:profile} and Figure \ref{fig:profiling} record the day-by-day distribution of the numbers of critical, high, medium, low and information-only severity for all assets.  Note that the number of vulnerabilities in most severity categories peaked around days 10-12 as that is the time when most of the assets were detailed to our team and VAPT was conducted on most assets. Also note that the number of vulnerabilities in various categories increased and decreased because some vendors fixed vulnerabilities in their products in a timely manner, but others did not.  Table \ref{tab:sla-plain-cells} records the exact duration that a vendor took to mitigate {\em critical vulnerabilities} in their respective assets. We show this in Table \ref{tab:sla-plain-cells} particularly to demonstrate how the vendors handled the VAPT team reports and how responsive they were to fixing critical severity vulnerabilities.  Unfortunately, some of the vendors did not mitigate all the vulnerabilities in their products even until the very start of the mega event.

\begin{table}[!htb]
    \centering
    \caption{Technical Vulnerability Distribution in Initial report}
    \label{tab:vapt}
    \begin{tabular}{p{7cm} c} \hline
    \textbf{ Vulnerability} & \textbf{Count} \\\hline
        Information disclosure & 7  \\
        Click-jacking & 4 \\
        Exposure of sensitive information & 3 \\
        No rate limiting & 3\\
        Enumeration of email/name/mobile & 2\\
        Vulnerable / outdated components & 2\\
        Cross site scripting & 2 \\
        Spoofing & 1 \\
        Unsigned certificate & 1 \\
        Cross domain referrer leakage & 1 \\
        Parameter tampering & 1 \\\hline
    \end{tabular}

\end{table}

\begin{table*}[!htb]
    \centering
    \caption{Profile of Technical Vulnerabilities}
    \label{tab:profile}
    {\small
    \setlength{\tabcolsep}{3pt}
    \begin{tabular}{l *{26}{c}}
        \toprule
        Severity $\downarrow$ & \multicolumn{26}{c}{Days $\rightarrow$} \\
        & $D_{1}$ & $D_{2}$ & $D_{3}$ & $D_{4}$ & $D_{5}$ & $D_{6}$ & $D_{7}$ & $D_{8}$ & $D_{9}$ & $D_{10}$ & $D_{11}$ & $D_{12}$ & $D_{13}$ & $D_{14}$ & $D_{15}$ & $D_{16}$ & $D_{17}$ & $D_{18}$ & $D_{19}$ & $D_{20}$ & $D_{21}$ & $D_{22}$ & $D_{23}$ & $D_{24}$ & $D_{25}$ & $D_{26}$ \\\hline
        \midrule
        Critical & 1  & 2  & 2  & 2  & 2  & 2  & 3  & 5  & 9  & 9  & 8  & 3  & 3  & 3  & 3  & 4  & 4  & 4  & 4  & 4  & 4  & 4  & 4  & 4  & 4  & 1  \\\hline
        CF & - & 1 & 2  & 2  & 2  & 2  & 2  & 3  & 5  & 9  & 9  & 8  & 3  & 3  & 3  & 3  & 4  & 4  & 4  & 4  & 4  & 4  & 4  & 4  & 4  & 4   \\
        NV & 1  & 1  & -  & -  & -  & - & 1  & 2  & 4  & -  & 1  & -  & -  & -  & 1  & 3  & -  & -  & -  & -  & -  & -  & -  & -  & -  & -  \\
        Mi & -  & -  & -  & -  & -  & -  & -  & -  & -  & -  & 2  & 5  & -  & -  & 1  & 2  & -  & -  & -  & -  & -  & -  & -  & -  & -  & 3  \\\hline \hline

        High & 6 & 8  & 9  & 9  & 13 & 13 & 20 & 24 & 28 & 28 & 26 & 16 & 14 & 11 & 14 & 14 & 14 & 14 & 10 & 10 & 10 & 10 & 10 & 10 & 13 & 13 \\\hline
        CF & - & 6 & 8  & 9  & 9  & 13 & 13 & 20 & 24 & 28 & 28 & 26 & 16 & 14 & 11 & 14 & 14 & 14 & 14 & 10 & 10 & 10 & 10 & 10 & 10 & 13   \\
        NV & 6  & 2  & 1  & -  & 8  & -  & 7  & 4  & 4  & -  & -  & -  & -  & -  & 3  & 1  & -  & -  & -  & -  & -  & -  & -  & -  & 3  & -  \\
        Mi & -  & -  & -  & -  & 4  & -  & -  & -  & -  & -  & 2  & 10  & 2  & 3  & -  & 1  & -  & -  & 4  & -  & -  & -  & -  & -  & -  & -  \\\hline \hline

        Medium & 14 & 20 & 21 & 21 & 22 & 22 & 23 & 26 & 26 & 26 & 27 & 16 & 15 & 11 & 13 & 16 & 16 & 16 & 14 & 14 & 14 & 15 & 14 & 14 & 15 & 14 \\\hline
        CF & - & 14 & 20 & 21 & 21 & 22 & 22 & 23 & 26 & 26 & 26 & 27 & 16 & 15 & 11 & 13 & 16 & 16 & 16 & 14 & 14 & 14 & 15 & 14 & 14 & 15   \\
        NV & 14  & 6  & 1  & -  & 5  & -  & 1  & 4  & 4  & -  & 4  & -  & -  & -  & 2  & 4  & -  & -  & -  & -  & -  & 1  & -  & -  & 1  & -  \\
        Mi & -  & -  & -  & -  & 4  & -  & -  & 1  & 4  & -  & 3  & 11  & 1  & 4  & -  & 1  & -  & -  & 2  & -  & -  & -  & 1  & -  & -  & 1  \\\hline \hline

        Low & 7  & 13 & 13 & 13 & 11 & 10 & 14 & 18 & 19 & 19 & 15 & 11 & 10 & 8  & 9  & 11 & 11  & 11  & 11  & 11  & 11  & 10  & 10  & 10  & 10  & 10  \\\hline
        CF &  - & 7  & 13 & 13 & 13 & 11 & 10 & 14 & 18 & 19 & 19 & 15 & 11 & 10 & 8  & 9  & 11 & 11  & 11  & 11  & 11  & 11  & 10  & 10  & 10  & 10   \\
        NV & 7  & 6  & -  & -  & 1  & -  & 4  & 4  & 1  & -  & 1  & -  & -  & - & 1  & 4  & -  & -  & -  & -  & -  & 1  & -  & -  & -  & -  \\
g        Mi & -  & -  & -  & -  & 3  & 1  & -  & -  & -  & -  & 5  & 4  & 1  & 2  & -  & 2  & -  & -  & -  & -  & -  & 2  & 0  & 0  & 0  & 0  \\\hline \hline

        Info & -  & -  & -  & -  & 1  & 1  & 1  & 1  & 1  & 1  & 1  & 1  & 1  & 1  & 1  & 1  & 1  & 1  & 1  & 1  & 1  & 1  & 1  & 1  & 1  & 1 \\\hline
        CF & -  & -  & -  & -  & -  & 1  & 1  & 1  & 1  & 1  & 1  & 1  & 1  & 1  & 1  & 1  & 1  & 1  & 1  & 1  & 1  & 1  & 1  & 1  & 1  & 1  \\
        NV & -  & -  & -  & -  & 1  & -  & -  & - & -  & -  & -  & -  & -  & -  & -  & -  & -  & -  & -  & -  & -  & -  & -  & -  & -  & -  \\
        Mi & -  & -  & -  & -  & -  & -  & -  & - & -  & -  & -  & -  & -  & -  & -  & -  & -  & -  & -  & -  & -  & -  & -  & -  & -  & -  \\\hline 
        \bottomrule
    \end{tabular}
    }
    \\[0.5em]
    Info: \hspace{0.2em}Informational; CF:\hspace{0.2em}Carry Forward; NV:\hspace{0.2em}New Vulnerability; Mi:\hspace{0.2em}Mitigated
\end{table*}

\begin{table*}[h]
\centering
\caption{SLA Metrics while responding to identified critical vulnerabilities}
\label{tab:sla-plain-cells}

\begin{tabular}{|c|c|*{26}{>{\centering\arraybackslash}p{0.12cm}|}c|}
\hline
\# & \textbf{Vulnerability} & \multicolumn{26}{c|}{\textbf{Day}} & \textbf{Resolution}\\
\cline{3-28}

 & $V_{i}a_{j}v_{k}$ $^{**}$ & \textbf{1} & \textbf{2} & \textbf{3} & \textbf{4} & \textbf{5} & \textbf{6} & \textbf{7} & \textbf{8} & \textbf{9} & \textbf{10}
 & \textbf{11} & \textbf{12} & \textbf{13} & \textbf{14} & \textbf{15} & \textbf{16} & \textbf{17} & \textbf{18} & \textbf{19} & \textbf{20}
 & \textbf{21} & \textbf{22} & \textbf{23} & \textbf{24} & \textbf{25} & \textbf{26} & Time (Days)\\
\hline

1  &    $V_{4}a_{1}v_{1}$     &         &         &         &         &         &         &         & \cellcolor{gray!40} & \cellcolor{gray!40} & \cellcolor{gray!40} & \cellcolor{gray!40} & \cellcolor{gray!40} & \cellcolor{gray!40} & \cellcolor{gray!40} & \cellcolor{gray!40} &         &         &         &         &         &         &         &         &         &         & & 8 \\\hline

2  &    $V_{4}a_{1}v_{2}$&                  &         &         &         &         &         &         & \cellcolor{gray!40} & \cellcolor{gray!40} & \cellcolor{gray!40} & \cellcolor{gray!40} & \cellcolor{gray!40} & \cellcolor{gray!40} & \cellcolor{gray!40} & \cellcolor{gray!40} &         &         &         &         &         &         &         &         &         &        & & 8 \\\hline

3 &    $V_{4}a_{2}v_{1}$         &         &         &         &         &         &         & \cellcolor{gray!40} & \cellcolor{gray!40} & \cellcolor{gray!40} & \cellcolor{gray!40} & \cellcolor{gray!40} & \cellcolor{gray!40} & \cellcolor{gray!40} & \cellcolor{gray!40} &         &         &         &         &         &         &         &         &         &         &&& 8 \\\hline

4 &    $V_{5}a_{1}v_{1}$ &         &         &         &         &         &         &         &         & \cellcolor{gray!40} & \cellcolor{gray!40} & \cellcolor{gray!40} &         &         &         &         &         &         &         &         &         &         &         &&&& & 3 \\\hline

5 &    $V_{5}a_{1}v_{2}$  &         &         &         &         &         &         &         &         & \cellcolor{gray!40} & \cellcolor{gray!40} & \cellcolor{gray!40} &         &         &         &         &         &         &         &         &         &         &         &&&&& 3 \\\hline

6 &    $V_{5}a_{1}v_{3}$  &         &         &         &         &         &         &         &         & \cellcolor{gray!40} & \cellcolor{gray!40} & \cellcolor{gray!40} &         &         &         &         &         &         &         &         &         &         &         &&&& & 3 \\\hline

7 &    $V_{5}a_{1}v_{4}$  &         &         &         &         &         &         &         &         & \cellcolor{gray!40} & \cellcolor{gray!40} &  &         &         &         &         &         &         &         &         &         &         &         &&&& & 2 \\\hline

8 &    $V_{5}a_{2}v_{1}$  &         &         &         &         &         &         &         &         &         &         & \cellcolor{gray!40} &         &         &         &         &         &         &         &         &         &         &         &         &         &         &         & 1 \\\hline
9 &    $V_{8}a_{1}v_{1}$  &         & \cellcolor{gray!40} & \cellcolor{gray!40} & \cellcolor{gray!40} & \cellcolor{gray!40} & \cellcolor{gray!40} & \cellcolor{gray!40} & \cellcolor{gray!40} & \cellcolor{gray!40} & \cellcolor{gray!40} & \cellcolor{gray!40} &         &         &         &         &         &         &         &         &         &         &         &         &         &         &         & 10 \\\hline
10 &    $V_{11}a_{1}v_{1}$ & \cellcolor{gray!40} & \cellcolor{gray!40} & \cellcolor{gray!40} & \cellcolor{gray!40} & \cellcolor{gray!40} & \cellcolor{gray!40} & \cellcolor{gray!40} & \cellcolor{gray!40} & \cellcolor{gray!40} & \cellcolor{gray!40} &         &         &         &         &         &         &         &         &         &         &         &         &         &         &         &         & 10 \\\hline
11 &    $V_{7}a_{1}v_{1}$  &         &         &         &         &         &         &         &         &         &         &         &         &         &         &  \cellcolor{gray!40}       & \cellcolor{gray!40} & \cellcolor{gray!40} & \cellcolor{gray!40} & \cellcolor{gray!40} & \cellcolor{gray!40} & \cellcolor{gray!40} & \cellcolor{gray!40} & \cellcolor{gray!40} & \cellcolor{gray!40} & \cellcolor{gray!40} & \cellcolor{gray!40} & 12 $^{*}$ \\\hline

12  &    $V_{12}a_{1}V_{1}$       &         &         &         &         &         &         &         &         &         &         &         &         &         &         &         & \cellcolor{gray!40}  & \cellcolor{gray!40}  & \cellcolor{gray!40} & \cellcolor{gray!40} & \cellcolor{gray!40} & \cellcolor{gray!40} & \cellcolor{gray!40} & \cellcolor{gray!40} & \cellcolor{gray!40} & \cellcolor{gray!40} & & 11 \\\hline

13  &    $V_{12}a_{1}V_{2}$       &         &         &         &         &         &         &         &         &         &         &         &         &         &         &         & \cellcolor{gray!40}  & \cellcolor{gray!40}  & \cellcolor{gray!40}  &  \cellcolor{gray!40}  & \cellcolor{gray!40} & \cellcolor{gray!40} & \cellcolor{gray!40} & \cellcolor{gray!40} & \cellcolor{gray!40} & \cellcolor{gray!40} & & 11 \\\hline

14   &    $V_{12}a_{1}V_{3}$      &         &         &         &         &         &         &         &         &         &         &         &         &         &         &         & \cellcolor{gray!40}  & \cellcolor{gray!40}  & \cellcolor{gray!40}   & \cellcolor{gray!40}  & \cellcolor{gray!40} & \cellcolor{gray!40} & \cellcolor{gray!40} & \cellcolor{gray!40} & \cellcolor{gray!40} & \cellcolor{gray!40} & & 11 \\\hline

\end{tabular}
    \\[0.5em]

$^{*}$\hspace{0.2em}No resolution considering 12 days\\
$^{**}$ $V_{i}a_{j}v_{k}$ is OEM enumeration with asset and vulnerability enumeration associated to the OEM 
\end{table*}

In addition to vulnerability assessment, another important task for risk assessment, even if  qualitative, is threat modeling for end-to-end processes, as well as threat modeling for individual applications. Examples of threat scenarios that we considered and communicated while making the qualitative risk assessment included: 
\begin{itemize}
    \item Disruption of  major applications by massive DDoS attacks
    \item Webpage defacement for landing pages of web applications, and various web pages for the event
    \item Exploitation of  applications to infect the backend servers with malware -- exfiltration of data, ransomware attack, wiper attack, drive-by-download, watering-hole attacks on individual users through the back-ends
    \item Exploitation of the web and mobile applications to bypass authentication, compromising personally identifiable data, data integrity violation
    \item Disruption of the  wired or wireless networks at the event site (network segments that connect CCTV surveillance, crowd control, public addressing systems, variable message display units, Command \& Control  etc. 
    \item Man-in-the-middle attack on CCTV camera feeds by  intercepting and replaying in order to hide some physical malicious activity  from the CCTV surveillance
    \item Publicizing fake applications, web sites, and fake service offers for potential visitors to the event defrauding them 
    \item Malicious and disinformation campaign on social media to generate panic and chaotic situation
    \item Inducing fake and/or offensive content on the variable message/video display boards used throughout the event site and outside creating embarrassment 
    \item Exploiting vulnerabilities in  applications such as reservation application, e-pass application, transportation reservation application, and others, mostly for committing financial frauds. 
    \item Attacking systems and applications at the ICCC (Integrated Command and Control Center) to reduce or compromise visibility to commit physical attacks and physical crimes on site.
    \item Insider attack from within the various ICT services and system integration vendors leading to disruption of services etc. 
    
\end{itemize}

Threat scenarios were evaluated against residual vulnerabilities and other mitigation measures, including existing on-ground police surveillance of suspicious activities, the difficulty for a hacker to set up a man-in-the-middle attack by ARP poisoning, or existing controls in the control center or the data centers hosting the applications. This step completed the high-level risk assessment. 

Once the high-level risk assessment was completed, risk-based control recommendations were made.  This step was inspired by the ISO/IEC 27001:2022 \cite{ISO27001} and ISO/IEC 27002:2022 \cite{ISO27002} standards. Due to time constraints, the applicability of all 93 controls listed in ISO / IEC 27002:2022 could not be formally evaluated against the ICT infrastructure and their roles in the business processes described above. Based on our prior knowledge of the ISO / IEC 27002: 2022 controls and NIST SP-800-53 \cite{NIST2020SP800-53r5},  our team developed a total of 252 checkpoints and provided the self-assessment checklist to each ICT vendor for self-reporting. Table \ref{tab:checklist} shows the distribution of the 252 controls between the four control categories (organizational, people, physical, and technological). Note that organizational controls and technological controls and corresponding checkpoints were by far the majority of controls and checkpoints in our suggested list.  We insist that these cyber security controls are established not only by the ICT infrastructure providers at the event site but also by all the vendors hosting the various web and mobile applications in their own or third-party data centers.

To summarize, beyond the high-level risk assessment, our approach to help organizers and their ICT vendors assess the adequacy of cyber security controls, we followed the four steps described below. Due to time pressure, some of these steps could not be fully documented, but they were roughly followed.

\begin{enumerate}

    \item {\bf Preparation and scoping}, where stakeholders establish scope and requirements:
    \begin{enumerate}
        \item {\bf Engage with the Event Organizers}
        \begin{itemize}
            \item Understand scope, business requirements, security targets, legal requirements, regulatory requirements, etc.
        \end{itemize}
        \item {\bf Coordinate with the third-party service providers}
        \begin{itemize}
            \item Understand their infrastructure as deployed for the event, their security posture, the security requirements imposed by the event authorities on them in terms of compliance. 
        \end{itemize}
        \item {\bf Develop a unified statement of applicability (SoA)} \end{enumerate}
    
    \item {\bf Log collection \& evidence management}, where a set of policies, procedures, logs, etc., are  collected \& expected from the vendor;
    \begin{enumerate}
        \item Collect cyber security policy, process documentation, their latest cyber audit reports (if any), their contracts with the event authority -- in particular, the sections relevant to cyber security requirements, etc.
        \item Collect Vulnerability Assessment and Penetration Testing (VAPT) and Information Security Management System (ISMS) reports or perform VAPT ourselves 
   \end{enumerate} 

    \item {\bf Gap analysis}, an iterative process with repeated evaluation of  the conformity against the controls \& associated checks, according to the counts mentioned in Table \ref{tab:checklist};
    \begin{enumerate}

        \item Assess organizational controls \& checkpoints
        \item Assess people controls \& checkpoints
        \item Assess physical controls \& checkpoints
        \item Assess technological controls \& checkpoints
    \end{enumerate}
    
    \item {\bf Recommendations based on the gap analysis }
    \begin{enumerate}

        \item Recommend patching instruction or mitigating control actions until re-assessment. 
        \item Reassess the vulnerabilities to verify closure or else continue suggesting mitigating controls, continue until residual risk is acceptable \end{enumerate}

\end{enumerate}

\begin{table}[ht]
    \centering
    \caption{Controls types and Checkpoints Count}
    \begin{tabular}{l c c}
        \toprule
        Category  & \makecell{No. Checkpoints \\(Supported Directly \& Indirectly)} \\
        \midrule
        Organizational & 80 \\
        People  & 16 \\
        Physical  & 22 \\
        Technological & 134 \\
        \bottomrule
    \end{tabular}
    \label{tab:checklist}
\end{table}

Finally, we developed a set of incident response playbooks complete with flow diagrams for creating incidence response playbooks. The following playbooks were particularly emphasized to the organizers for carrying out planning and pre-event exercise or tabletop exercises.  
\begin{itemize}
    \item Ransomware Incident
    \item DDoS
    \item Phishing
    \item Social Engineering
    \item Insider Threat
    \item Malware Attack
    \item Web Application Security Compromise 
    \item Physical Attack on cyber infrastructure
\end{itemize}

Although, at the start of the event, as our engagement wore off throughout the Mahakumbh event, our team was prepared to be on call for incident response or recovery assistance. However, serendipitously, no successful cyber attack took place during the 45 days of the event. Although our team made an unwavering effort to provide cyber resiliency following the NIST cyber resiliency engineering processes, and followed standard practices for high-level risk assessment through threat modeling, vulnerability assessment, and risk-based control recommendations, there were other teams such as the Indian CERT-IN (Computer Emergency Response Team of India), NCIIPC (National Critical Information Infrastructure Protection Center) also looked into vulnerabilities and other cyber security posture-related arrangements. Therefore, the incident-free mega event was possible through the effort of various stake holders and our part has been to follow a systematic methodology and also to learn from this exercise so we can do a better job during the next mega event. 

The following recommendations were also made to the organizers:

\subsection{Recommendations}

We requested that all ICT vendors, including the system integrator in charge of the overall digital infrastructure of the event, and all vendors developing and/or hosting applications undergo a set of measures and report back their findings. Among the measures requested, the following are examples of some of the important ones: 

\textbf{Procedural Measures}
\begin{itemize}
    \item \textbf{Documentation Reviews:} Make sure that the policies for access control, incident response, and data handling are documented and reviewed before the start of the event to align with the checklist of controls provided, ensuring clarity for auditors and staff.

    \item \textbf{Statement of Applicability (SoA) Mapping:} The SoA outlines which of the 252 controls apply, why others are excluded, and how they address identified risks, critical to audit trails. The SoA must accompany the reported findings. 

    \item \textbf{Training and Awareness:} Temporary staff must receive targeted training in the recognition of phishing and other social engineering tricks, and how to practice proper cyber hygiene. 

    \item \textbf{Incident Response Drills:} Simulated breaches (e.g. ransomware on a ticketing server) should be performed multiple times before the start of the event to test response times and coordination among IT, security and law enforcement teams. This would also help them get familiar with the incident response playbooks designed. 
\end{itemize}

\textbf{Technical Measures}
\begin{itemize}
    \item \textbf{MAC-Port Binding:} Devices such as staff laptops and ticketing kiosks supporting the E-pass system (public access through the Internet, private access for administrative functions) must be bound to specific switch ports through MAC addresses. This restricts rogue device access in exposed, temporary network environments.

    \item \textbf{ARP Spoofing Tests:} Address Resolution Protocol (ARP) spoofing can enable man-in-the-middle attacks. Regular testing with tools like \textit{Wireshark} or custom scripts must be used to detect anomalies, ensuring the integrity of the network layer.

    \item \textbf{Network Intrusion Detection Systems (NIDS):} Deploying NIDS (e.g., \textit{Snort} or \textit{Suricata}) provides real-time traffic monitoring for signatures of known attacks (e.g. SQL injection) and behavioral anomalies (e.g. sudden traffic spikes indicative of DDoS). 

    \item \textbf{Endpoint Encryption:} All data on endpoints - whether stored or in transit - must use AES-256 encryption, protecting sensitive information such as attendee details or payment records against physical theft or interception.

    \item \textbf{DDoS Mitigation:} Cloud-based services like \textit{Cloudflare} or \textit{Akamai} must be used to filter malicious traffic, ensuring the availability of critical services under high load.
\end{itemize}

Table \ref{tab:findings} shows how the 12 vendors evolved their controls over the one-month period through three assessment cycles. Since they were hosting the backend infrastructure for the applications on their servers or data centers, it was particularly important that they had appropriate controls to protect the backend servers as well as the collected data. 

\begin{table}[htbp]
\caption{Assessment and Conformity Evaluation}
\centering
\renewcommand\cellalign{lc}
\begin{tabular}{p{1cm} p{1.5cm} p{1.5cm} p{1cm} | c}
\toprule
\textbf{Vendor} & \multicolumn{3}{c|}{\textbf{Assessment}} & \textbf{Conformity (\%)} \\
                & 1\textsuperscript{st} & 2\textsuperscript{nd} & 3\textsuperscript{rd} & Out of 252 \\
\midrule
$V_{1}$  & 05 & 72 & 214 & 84.92\% \\
$V_{2}$  & 0 & 191 & 191 & 75.79\% \\
$V_{3}$  & 0 & 209 & 222 & 88.09\% \\
$V_{4}$  & 0   & 0   & 0 & 0\%  \\
$V_{5}$  & 0   & 0 & 42 & 16.66\% \\
$V_{6}$  & 0   & 0 & 41 & 16.26\% \\
$V_{7}$  & 0   & 0 & 15 & 5.95\%  \\
$V_{8}$  & 0   & 0 & 178 & 70.63\% \\
$V_{9}$  & 0   & 0 & 113 & 44.84\% \\
$V_{10}$ & 0   & 0 & 147 & 58.33\% \\
$V_{11}$ & 0   & 0 & 177 & 70.23\% \\
$V_{12}$ & 0   & 0 & 233 & 92.46\% \\
\bottomrule
\end{tabular}
\label{tab:findings}
\end{table}

\begin{table}[!htb]
    \centering
    \caption{Technical Vulnerability Distribution}
    \label{tab:vaptdistribution}
    \begin{tabular}{p{3cm}|p{3cm}} \hline
        Critical & 14  \\
        High & 36 \\
        Medium & 48 \\
        Low & 31\\
        Informational & 1\\
        Best Practices & 0\\ \hline
        Total & 130 \\\hline
    \end{tabular}

\end{table}

\section{Conclusion \& Future Work}
Our team got into action about 40 days before the start date of the event. The infrastructure was laid at the event venue during this period and applications were being developed. A system integration vendor that coordinated with multiple vendors and our team was used.

In the future, our recommendation would be to involve the subject matter expert team much earlier than the start date of the event. The cyber resilience engineering process must be planned from the very beginning of the event planning which should precede by at least one year from the start date. 

Future event organizers should also perform asset inventory, perform a detailed risk assessment, and implement risk-based controls according to a standard for ISMS such as the ISO 27001 series of standards.   

Finally, during the event, the command and control center must also host an industrial strength SoC. Continuous threat-hunting, and incident response must be a default activity for all mega events. Threats observed must be compiled, and communicated, and analyzed to validate if the controls in place were effective, and the scopes of improvement in the future.

In summary, while our team along with the organizers have been fortunate enough to not suffer any major cyber attack, and the entire duration of the mega event went by without any report worthy cyber incident, we believe that we can improve the entire process, methodology, and outcome based on the learning we achieved this time. This paper is a summary of our methodology, and learning, which we believe could help the organizers of similar mega events in the future.  

\section*{Acknowledgment}
The authors would like to sincerely thank Shubham Khanduri, Shubham Malviya and Other memebers from VAPT team for their valuable support, technical assistance, and helpful discussions. Their contributions were instrumental in the successful completion of this study.


\end{document}